\documentclass[onecolumn,aps,amsmath,amssymb]{revtex4}

\usepackage{graphicx}
\usepackage{dcolumn}
\usepackage{bm}

\newcommand{\be}{\begin{eqnarray}}
\newcommand{\ee}{\end{eqnarray}}

\begin{document}
\title{Higgs Mechanism  via  Bose-Einstein Condensation}

 \author{Francesco {\sc Sannino}}
 \email{francesco.sannino@nbi.dk}
 \author{Kimmo {\sc Tuominen}}\email{tuominen@nordita.dk}
 \affiliation{NORDITA, Blegdamsvej 17,
 DK-2100 Copenhagen \O, Denmark }
\date{June 2003}

\begin{abstract}
Recently the Bose-Einstein phenomenon has been proposed as
possible physical mechanism underlying the spontaneous symmetry
breaking in cold gauge theories. The mechanism is natural and we
use it to drive the electroweak symmetry breaking. The mechanism
can be implemented in different ways while here we review two
simple models in which the Bose-Einstein sector is felt directly
or indirectly by all of the standard model fields. The structure
of the corrections due to the new mechanism is general and
independent on the model used leading to experimental signatures
which can be disentangled from other known extensions of the
standard model.
\end{abstract}

\maketitle


\section{Introduction}

The Standard Model of particle interactions has passed numerous
experimental tests \cite{Hagiwara:fs} but despite the experimental
successes it is commonly believed that it is still incomplete. The
spontaneous breaking of the electroweak symmetry, for example, has
been the subject of intensive studies and different models have
been proposed to try to accommodate it in a more general and
satisfactory framework. Technicolor theories \cite{Hill:2002ap}
and supersymmetric extensions \cite{Kane:2002tr} of the Standard
Model are two relevant examples.

In \cite{Sannino:2003mt,{Sannino:2003ai}} we explored the
possibility that the relativistic Bose-Einstein phenomenon
\cite{Haber:1981ts,Kapusta:aa} is the cause of electroweak
symmetry breaking. In \cite{Sannino:2003mt} we introduced a new
global symmetry of the Higgs field and associated a chemical
potential $\mu$ with the generators of such a new symmetry. A
relevant property of the theory was that the chemical potential
induced a, non renormalizable and hence protected against
quadratic divergences, direct negative mass squared for the Higgs
field at the tree level destabilizing the symmetric vacuum and
triggering symmetry breaking. The local gauge symmetries were
broken spontaneously and the associated gauge bosons acquired a
standard mass term. We also showed that while the properties of
the massive gauge bosons at the tree level are identical to the
ones induced by the conventional Higgs mechanism, the Higgs field
itself has specific Lorentz non covariant dispersion relations.
The Bose-Einstein mechanism occurs, in fact, in a specific frame
the effects of which are felt by the other particles in the theory
via electroweak radiative corrections. Some of these corrections
have been explicitly computed in \cite{Sannino:2003mt} and the
model presented in \cite{Sannino:2003mt} is very predictive.

\begin{figure}[bth]
 \includegraphics[width=8.5truecm, clip=true]{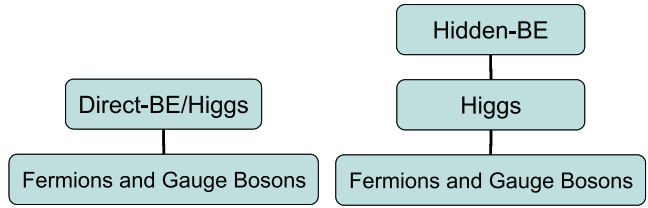}
 \caption{Schematic representation of the direct and indirect
 Bose-Einstein mechanism for triggering electroweak symmetry
 breaking.}
 \label{Figura1}
\end{figure}

In \cite{Sannino:2003ai} we presented a new model in which a
hidden Bose-Einstein sector drives the electroweak symmetry
breaking while yielding small corrections to the standard Higgs
dispersion relations. In this way the presence of a frame and
hence Lorentz breaking corrections are suppressed with respect to
the ones shown in \cite{Sannino:2003mt}. In this class of models
the Bose-Einstein mechanism operates on a complex scalar field
neutral under all of the Standard Model interactions. On general
grounds this field interacts with the ordinary Higgs and we used
these interactions to trigger the ordinary electroweak symmetry
breaking. Due to the Bose-Einstein nature of the new mechanism the
form of the corrections is insensitive to its specific
realization, hence the mechanism is distinguishable from other
sources of beyond standard model physics. The two classes of
models we presented in \cite{Sannino:2003mt,{Sannino:2003ai}} are
schematically summarized in figure \ref{Figura1}. In the first one
the Higgs field feels directly the presence of a net background
charge and communicates it to the gauge bosons and fermions via
higher order corrections (left panel). In the second (the hidden
case) a singlet field with respect to the standard model quantum
numbers directly feels the effects of a net background charge. The
ordinary Higgs field weakly feels the effects of a frame via the
interactions with the singlet field. Finally the rest of the
standard model particles will be affected via higher order
corrections (right panel).
 Since the chemical potential differentiates between
time and space, and we have a scalar vacuum, all of the dispersion
relations are isotropic in the Bose-Einstein frame. Theories with
condensates of vectorial type have been studied in different
realms of theoretical physics
\cite{Linde:1979pr,{Kapusta:1990qc},Ferrer:jc,Ferrer:ag,{Ambjorn:1989gb},{Kajantie:1998rz},{Sannino:2002wp}}.
It is also worth mentioning that, although in a different
framework, effects of a large lepton number on the spontaneous
gauge symmetry breaking for the electroweak theory at high
temperature relevant for the early universe have been studied
\cite{Linde:1979pr,{Ferrer:jc},{Ferrer:ag},{Bajc:1999he},{Bajc:1997ky}}.
In \cite{Mazur:2001fv} the concept of Bose-Einstein condensation
in gravitational systems was considered.

\section{Model I: Direct Bose-Einstein Mechanism}\label{tre}
The Higgs sector of the Standard Model
possesses, when the gauge couplings are switched off, an
$SU_L(2)\times SU_R(2)$ symmetry. The full symmetry group is
mostly easily recognized when the Higgs
doublet field \begin{eqnarray} H=\frac{1}{\sqrt{2}}\left(%
\begin{array}{c}
  \pi_2 + i\, \pi_1 \\
  \sigma - i\, \pi_3 \\
\end{array}%
\right)\end{eqnarray} is represented as a two by two matrix in the
following way:
\begin{eqnarray}
M\equiv\frac{1}{\sqrt{2}}\left(\sigma + i\,
\vec{\tau}\cdot\vec{\pi} \right)=\frac{1}{\sqrt{2}}\left(%
\begin{array}{cc}
   \sigma + i\, \pi_3 & \pi_2 + i\, \pi_1  \\
  -\pi_2 + i\, \pi_1 & \sigma - i\, \pi_3 \\
\end{array}%
\right) \equiv\left[i\,\tau_2H^{\ast}\ ,\,H\right]
 \ , \label{notation}
\end{eqnarray}
where the two columns of the two by two matrix are identified with
$i\tau_2 H^{\ast}$ and $H$. The $SU_L(2)\times SU_R(2)$ group acts
linearly on $M$ according to:
\begin{eqnarray}
M\rightarrow g_L M g_R^{\dagger} \qquad {\rm and} \qquad g_{L/R}
\in SU_{L/R}(2)\ .
\end{eqnarray}
It is easy to verify that:
\begin{eqnarray}
M\frac{\left(1-\tau^3\right)}{2} = \left[0\ , \, H\right] \ ,
\qquad M\frac{\left(1+\tau^3\right)}{2} = \left[i\,\tau_2H^{\ast}
\ , \, 0\right] \ ,
\end{eqnarray}
where the notation follows from (\ref{notation}) and the zeros
represent an entire column of the two by two matrix. The $SU_L(2)$
symmetry is gauged by introducing the weak gauge bosons $W^a$ with
$a=1,2,3$. The hypercharge generator is taken to be the third
generator of $SU_R(2)$. The ordinary covariant derivative acting
on the Higgs, in the present notation, is:
\begin{eqnarray}
D_{\mu}M=\partial_{\mu}M -i\,g\,W_{\mu}M + i\,g^{\prime}M\,B_{\mu}
\ , \qquad {\rm with}\qquad W_{\mu}=W_{\mu}^{a}\frac{\tau^{a}}{2}
\ ,\quad B_{\mu}=B_{\mu}\frac{\tau^{3}}{2} \ .
\end{eqnarray}
At this point one simply {\it assumes} that the mass squared of
the Higgs field is negative and this leads to the electroweak
symmetry breaking and more generally to the successful Standard
Model as we know. However, theoretically a more satisfactory
explanation of the origin of the Higgs mechanism is needed. In the
literature many models have been proposed in order to explain the
emergence of such a negative mass squared. Technicolor theories,
for example, assume a dynamical mechanism identical to spontaneous
chiral symmetry breaking in quantum chromodynamics
\cite{Hill:2002ap}. Supersymmetric extensions of the Standard
Model \cite{Kane:2002tr} explain the negative mass squared as due
to the running of the masses from high scales down to the
electroweak one.

We now review the first model in which the Higgs mechanism is
driven by the Bose-Einstein phenomenon \cite{Sannino:2003mt}. To
illustrate the idea we consider a Higgs sector with the symmetry
group $SU_L(2)\times SU_R(2)\times U_A(1)$ where the $SU_L(2)$ is
later on gauged and the $U_Y(1)$ is associated to the
$T^3=\frac{\tau^3}{2}$ generator of $SU_R(2)$ while $U_A(1)$
remains a global symmetry. Now we introduce a chemical potential
$\mu_A$ associated to $U_A(1)$. When the chemical potential is
sufficiently large $SU_L(2)\times SU_R(2)\times U_A(1)$ breaks
spontaneously to $SU_V(2)$ and we have four goldstones. It is
advantageous to use nonlinear realizations for the Higgs field
with:
\begin{eqnarray}
M=\frac{\sigma}{\sqrt{2}}\,U_{\eta}\,U \qquad {\rm with} \qquad
U_{\eta}=e^{i\frac{\eta}{v}} \,\qquad U=e^{i\frac{\pi}{v}}\ ,
\qquad {\rm and } \qquad \pi={\tau^a}\pi^a \ .
\end{eqnarray}
In the above equation $v$ is the vacuum expectation value of
$\sigma$. In the linearly realized case we should also include the
heavy $U_A(1)$ partners of the $\pi$ field which we have taken to
be more massive than the neutral Higgs particle and hence we have
decoupled them. The $\eta$ field is the the goldstone boson
associated to the spontaneous breaking of the global $U_A(1)$
symmetry.

The gauge interactions as well as the chemical potential are
introduced via the following covariant derivative:
\begin{eqnarray}
{\cal D}_{\mu}M=\partial_{\mu}M - i\,g\,W_{\mu}\,M +
i\,g^{\prime}\,M\,B_{\mu}-i{\cal A}_{\mu}M \equiv {D}_{\mu}M -
i{\cal A}_{\mu}M\ , \quad {\rm with} \quad {\cal
A}_{\mu}={\mu_{A}} (1,\vec{0}) \ .
\end{eqnarray}
Electroweak breaking is now forced by the introduction of the
chemical potential for the extra global symmetry.
{}For ${\mu_A^2}>m^2$ we have Bose-Einstein condensation together
with the ordinary spontaneous breaking of the internal symmetry
$SU_L(2)\times SU_R(2)\times U_A(1)\rightarrow SU_V(2)$  with 4
null curvatures corresponding to the four broken generators. In
the unitary gauge the three fields $\pi^a$ are absorbed into the
longitudinal components of the three massive gauge boson fields
while the field $\eta$ remains massless. In the unitary gauge the
quadratic terms are:
\begin{eqnarray}
{\cal L}_{\rm quadratic}&=&\frac{1}{2} \partial_{\mu}h
\partial^{\mu}h
+\frac{1}{2}\partial_{\mu}{\eta}\partial^{\mu}{\eta}  - \mu_A
\left(h\partial_0 \eta - \eta
\partial_0 h\right) \nonumber \\&+&\frac{v^2}{8}\, \left[g^2\,\left(W_{\mu}^1
W^{\mu,1} +W_{\mu}^2 W^{\mu,2}\right)+ \left(g\,W_{\mu}^3 -
g^{\prime}\,B_{\mu}\right)^2\right] -(\mu^2_A -{m^2})h^2
\end{eqnarray}
with
\begin{eqnarray}\langle \sigma \rangle^2 =
v^2={\frac{{\mu_A^2}-m^2}{\lambda}}\ , \qquad {\rm and} \qquad
\sigma = v + h \ ,
\end{eqnarray}
where $h$ is the Higgs field. The $Z_{\mu}$ and the photon
$A_{\mu}$ gauge bosons are:
\begin{eqnarray}
Z_{\mu}=\cos\theta_W\, W_{\mu}^3 - \sin\theta_{W}B_{\mu} \ ,\qquad
A_{\mu}=\cos\theta_W\, B_{\mu} + \sin\theta_{W}W_{\mu}^3 \ ,
\end{eqnarray}
with $\tan\theta_{W}=g^{\prime}/g$ while the charged massive
vector bosons are $W^{\pm}_{\mu}=(W^1\pm i\,W^2_{\mu})/\sqrt{2}$.
The bosons masses, due to the custodial symmetry, satisfy the tree
level relation $M^2_Z=M^2_W/\cos^2\theta_{W}$ with
$M^2_W=g^2\,v^2/4$.

Except for the presence of an extra degree of freedom and the
$h-\eta$ mixing term one recovers the correct electroweak symmetry
breaking pattern. The third term in the Lagrangian signals an
explicit breaking of the Lorentz symmetry in the Higgs sector.
Such a breaking is due to the introduction of the chemical
potential and hence it happens in a very specific and predictive
way so that all of the features can be studied. Note that in our
theory Lorentz breaking, at the tree level, is confined only to
the Higgs sector of the theory which is also the least known
experimentally. The rest of the theory is affected via weak
radiative corrections. As previously emphasized, in general, the
introduction of the chemical potential at zero temperature does
not introduce new ultraviolet divergences and hence does not spoil
renormalizability. Besides, diagonalizing the quadratic terms the
spectrum and the propagators for $h$ and $\eta$ yields:
\begin{eqnarray}
E^2_{h}&=& \Delta^2 + \left(1 +
4\frac{\mu_{A}}{\Delta^2}\right)p^2 - \frac{\mu_{A}^4}{\Delta^6}
p^4 + \cdots  \ , \\
E^2_{\eta}&=&\left(1 - 4\frac{\mu_{A}^2}{\Delta^2}\right)p^2 +
\cdots.
\end{eqnarray}
with
\begin{eqnarray}
\Delta^2 = 2(3\mu^2_{A} - m^2)=4\mu^2_{A} + 2\,\left({\mu^2_{A}} -
m^2\right) \ .
\end{eqnarray}
The second term in the expression for $\Delta^2$ is the potential
curvature evaluated on the ground state which in the absence of
the chemical potential is the mass of $h$. Note that the energy
gap (energy at zero momentum) of the Higgs $\Delta$ is larger than
the one predicted by just assuming a change in the sign of the
mass squared coefficient.

Some important features of the present Bose-Einstein condensation
mechanism are: a mass $\Delta$ for $h$ larger than the one in
vacuum i.e. $M_h^2=2\left(\mu^2_A - m^2\right)$, more specifically
$\Delta \geq \sqrt{3} M_h$, and a massless goldstone state $\eta$.
{}For example if the Higgs mass, in vacuum, is about $90$~GeV the
Bose-Einstein mechanism leads to a mass larger than $155$~GeV. The
spontaneous breaking of the $U_A(1)$ symmetry requires the $\eta$
field to be massless. Phenomenologically one may need to add a
mass, small with respect to the chemical potential, for the $\eta$
field \cite{Sannino:2003mt}.

\section{The non Higgs Sector}\label{quattro}
\begin{figure}[hbt]
 \includegraphics[width=9truecm, clip=true]{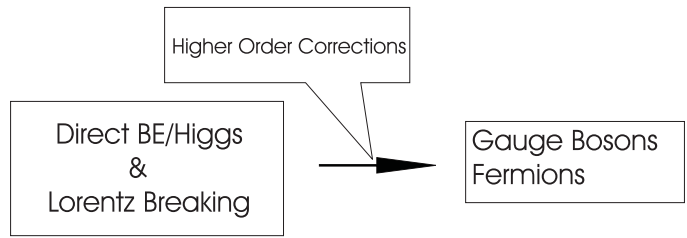}
 \caption{Schematic representation of the way the corrections due to the direct
 Bose-Einstein condensation are felt by the non Higgs sector of the standard model.}
 \label{Figura2}
 \end{figure}

We have shown that, at the tree level, the gauge bosons acquire
the ordinary electroweak masses and dispersion relations while we
argued that deviations with respect to the ordinary Higgs
mechanism, in our theory, arise when considering higher order
corrections. In this section we analyze some of the effects of
spontaneous symmetry breaking via a nonzero $U_A(1)$ charge
density on the non Higgs sector of the electroweak theory due to
such higher order effects. We first investigate the gauge boson
sector and then the fermion one. On general grounds we expect the
presence of the chemical potential to induce different time and
spatial corrections while keeping rotational invariance intact. In
Fig.~\ref{Figura2} we schematically show how the corrections due
to the Bose-Einstein induced Higgs mechanism propagate to the rest
of the standard model fields.
\subsection{The Gauge Bosons}
The Higgs propagator is modified in the presence of the chemical
potential and assumes the form:
\begin{eqnarray}
\parbox{2cm}{\includegraphics[width=20cm,clip=true]{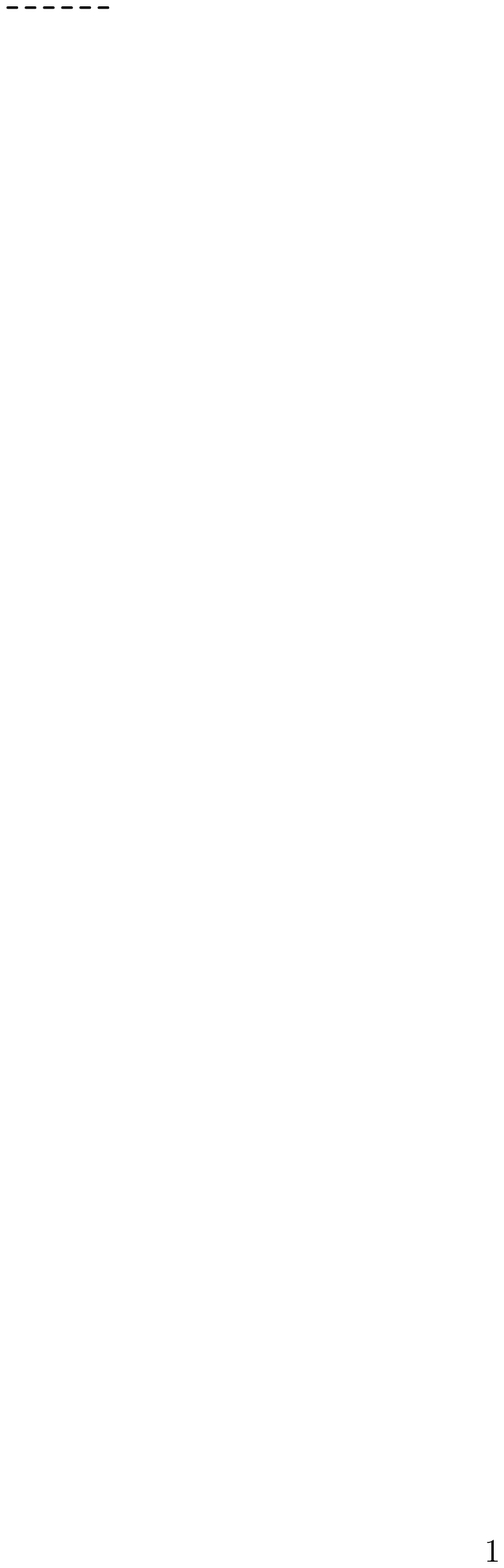}}
&=& i\,\frac{p^2}{p^4-2(\mu^2_A-m^2)p^2-4\mu^2_Ap_0^2} \ .
\end{eqnarray}
All of the loops containing this propagator are affected by the
presence of the chemical potential. The Landau gauge
\cite{Cheng-Li} is chosen to evaluate the relevant contributions
although our results are gauge independent.

We are interested in computing the new physics corrections for the
$W$ vacuum polarization due to a different Higgs sector with
respect to the conventional Standard Model one. The diagrams
needed are \cite{Peskin:1991sw,Takeuchi:1992bu}:
\begin{equation}
\parbox{9cm}{\includegraphics[width=9cm,clip=true]{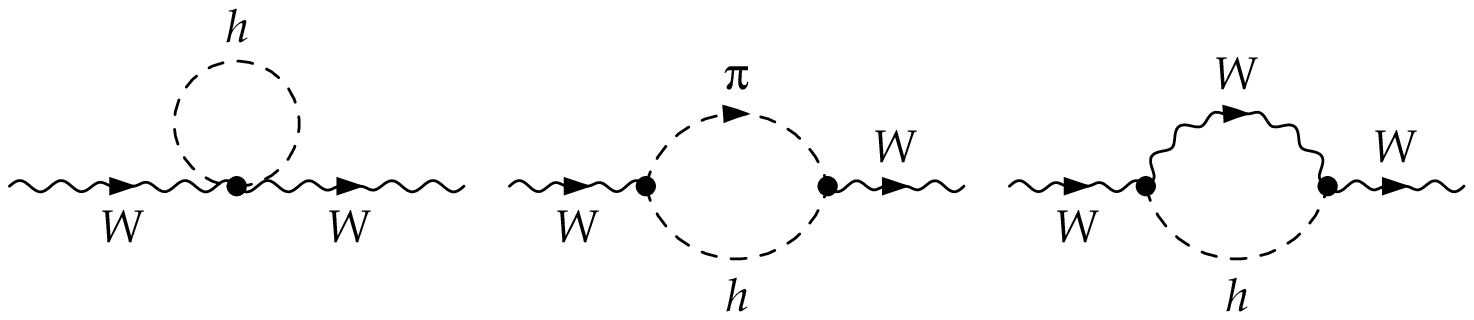}} \label{WZ-Self-Energies}
\end{equation}
The major difference with respect to known possible extensions is
the appearance of a new type of dispersion relations for the
Higgs. The $\eta$ particle does not appear in the previous
diagrams since we used a polar decomposition for the $M$ field.

In the leading logarithmic approximation, when the external
momentum vanishes and assuming an expansion in the gauge bosons
masses with respect to the Higgs mass and setting $m=0$, the
result is:
\begin{eqnarray}
\Pi_{WW}^{\mu\nu}(0)&=&
\frac{3\,g^2}{4\,(4\pi)^2}\,\log\left(\frac{\mu^2_A(2+\sqrt{3})}{\Lambda^2}\right)
\left[g_{\mu\nu} \,\left( M^2_W +
\frac{\mu^2_A}{9}\right)-\frac{4\,\mu^2_A}{9}\,V_{\mu}V_{\nu}
\right]\nonumber \\ &+& {\cal
O}\left(\frac{M^2_{W}}{\mu^2_A}\right)
\end{eqnarray}
where $\Lambda$ is the renormalization scale and
$V_{\mu}=\left(1,\bf{0}\right)$. The photon vacuum polarization is
not affected at zero momentum due to the ordinary Ward identities
\cite{Takeuchi:1992bu}. The first diagram on the right hand side
of eq.~(\ref{WZ-Self-Energies}) does not contribute to the Lorentz
non covariance of the vacuum polarization which is entirely due to
the second and third diagrams. Note that the specific combination
$\mu_A^2\left(2+\sqrt{3}\right)$ appearing in any logarithmic
corrections is consequence of the fact that in the presence of the
chemical potential the particle gaps are not the curvatures
evaluated on the minimum.

The contribution to the $Z$ vacuum polarization is obtained by
replacing in the previous expressions $g^2$ with
$g^2/\cos^2\theta_W$ and $M^2_W$ with $M^2_Z$. The  onset of
Lorentz breaking in this sector is small especially if one chooses
$\Lambda$ of the order of $\mu_A$. In general it is possible to
define a new set of oblique parameters capable to capture the
relevant corrections due to this type of spontaneous symmetry
breaking. Here, for illustration, we consider the following
straightforward extension of the parameter which measures
deviations with respect to the breaking of the custodial symmetry
i.e. the parameter $T$ \cite{Peskin:1991sw}:
\begin{eqnarray}
\alpha\,T^{\mu\nu}=\frac{e^2}{\sin^2\theta_W\,\cos^2_W\,M^2_Z}
\left({\Pi_{11}^{\mu\nu}}^{,~new}(0)-{\Pi_{33}^{\mu\nu}}^{,~new}(0)\right)
\ ,\end{eqnarray} with $\alpha=e^2/4\pi$ the fine structure
constant. We also used
$g^2\,\Pi_{11}^{\mu\nu}(0)=\Pi_{WW}^{\mu\nu}(0)$ and
${g^2}\Pi_{33}^{\mu\nu}(0)=\cos^2\theta_W\Pi_{ZZ}^{\mu\nu}(0)$
since the photon vacuum polarization at zero momentum vanishes.

This parameter is equal to $\alpha\,T\,g^{\mu\nu}$ for any Lorentz
preserving extension of the Higgs sector. The newly defined
parameter is not directly a measure of the amount of Lorentz
breaking but rather it estimates the amount of custodial symmetry
breaking for the different spacetime components of the vacuum
polarizations. Here we still have $T^{\mu\nu}=T\,g^{\mu\nu}$ but
due to the fact that we have different dispersion relations and a
different gap structure with the respect to the Standard Model
masses the value of $T$, although very small, is not zero:
\begin{eqnarray}
T\approx-\frac{3}{16\pi}\frac{1}{\cos^2\theta_W}\log\left(1+\frac{\sqrt{3}}{2}\right)\approx
-0.048 \ .
\end{eqnarray}
Here we assumed the Standard Model Higgs mass to be given by the
expression $M^2_H=2\mu_A^2$ which is the curvature evaluated on
the minimum.
\subsection{The Fermions}
The fermions constitute a very interesting sector to be explored
since it can be used experimentally to test the idea presented in
this paper. In order to understand the type of corrections we
start with recalling that the chemical potential explicitly breaks
$SL(2,C)$ to $SO(3)$. So the corrections must differentiate time
from space in the fermion kinetic term according to:
\begin{eqnarray}
(1 - a_0)\,\bar{f}\gamma^0\partial_0\, f+ (1-a)
\,\bar{f}\gamma^{i}\partial_i\, f\rightarrow
\bar{f}\gamma^0\partial_0\, f + v_{f}\,
\bar{f}\gamma^i\partial_i\, f \ ,\qquad  v_{f}\simeq 1-(a-a_0)
\label{fermion-dr}
\end{eqnarray}
where $a$ and $a_0$ are the corrections induced by loop
contributions and $f$ represents a generic Standard Model fermion.
In the last expression we have rescaled the fermion wave function
and used the fact that the $a$'s are small calculable corrections.
In the difference $a-a_0$ all of the Lorentz covariant corrections
disappear while only the Lorentz breaking terms survive.

In order for the fermions to receive one loop corrections
sensitive to the chemical potential they need to couple at the
tree level with the Higgs. This is achieved via the Yukawa
interactions (for a more detailed discussion see
\cite{Sannino:2003mt}):
\begin{eqnarray}
\widetilde{Y}_{f}\, h \bar{f}f \ , \quad {\rm with} \quad
\widetilde{Y}_{f}\simeq \frac{m_{f}}{v} \ .
\end{eqnarray}
To determine $v_{f}$ at the one loop we need to compute the
contributions to the fermion self energy which break Lorentz
invariance. In the present case the one loop diagram is
\begin{equation}
\parbox{3.5cm}{\includegraphics[width=3.5cm,clip=true]{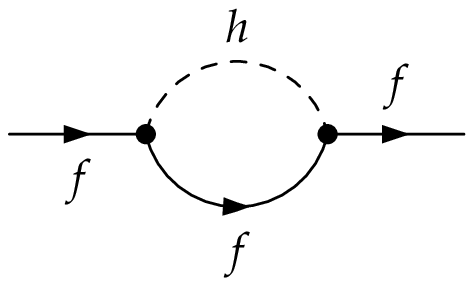}},
 \label{ftext_integral}
\end{equation}
where the solid line represent the fermion and the dashed one the
Higgs field. The diagram is evaluated in detail in
\cite{Sannino:2003mt} and yields the following contribution to the
$a$ and $a_0$ coefficients:
\begin{eqnarray}
a&=&\frac{\widetilde{Y}_{f}^2}{2(4 \pi)^2} \left[\log
\left(\frac{\mu^2_A(2+\sqrt{3})}{\Lambda^2}\right)-
\frac{1}{6}\right] \ , \\
a_0&=&\frac{\widetilde{Y}_{f}^2}{2(4 \pi)^2}
\left[\log\left(\frac{\mu^2_A(2+\sqrt{3})}{\Lambda^2}\right)+4\sqrt{3}
-\frac{15}{2}\right] \ .
\end{eqnarray}
If only the leading logarithmic corrections are kept we have that
the fermions still obey standard Lorentz covariant dispersion
relations. The dispersion relations are modified when considering
the finite contributions. To estimate the size of possible
departures from the standard dispersion relations we keep the
constant terms and determine
\begin{eqnarray}
v_{e}\simeq 1 - \widetilde{Y}_e^2\frac{0.4}{2(4\pi)^2} \simeq 1 -
5\times 10^{-15} \ ,
\end{eqnarray}
where the numerical evaluation has been performed for the
electron. The present formula is valid practically for all of the
fermions. For the muon the corrections are of the order of
$8\times 10^{-10}$. The induced corrections for the photon due to
fermion loops are further suppressed by powers of the fine
structure constant $\alpha=e^2/4\pi$.

\section{Modified Fermi Theory}
Another class of indirect corrections to the fermion
sector are the ones induced by modified gauge vector propagators
discussed in the previous section. To illustrate these effects we
use the low energy electroweak effective theory for the charged
currents. The neutral currents will be affected in a similar way.
The chemical potential leaves intact the rotational subgroups of
the Lorentz transformations, so the effective Lagrangian modifies
as follows:
\begin{eqnarray}
{\cal L}^{\rm CC }_{\rm Eff}=-2\sqrt{2}\,G
J_{\mu}^{+}{J^{\mu}}^{-} \Rightarrow
-2\sqrt{2}\,G\,\left[{J_{\mu}}^{+}{J^{\mu}}^{-}+ \delta\,
J_{i}^{+}{J^{i}}^{-}\right] \ , \label{fermitheory}
\end{eqnarray}
where $\delta$ is a coefficient effectively measuring the
corrections due to modified dispersion relations for the gauge
vectors. Using the previous Lagrangian the decay rate for the
process $\mu\rightarrow e \bar{\nu}_{e}\nu_{\mu}$ is:
\begin{eqnarray}
\Gamma[\mu\rightarrow e \bar{\nu}_{e}\nu_{\mu}]=\frac{G^2
M_{\mu}^5}{192\pi^3}\left(1 + \frac{3}{2}\delta\right) \ ,
\label{muondec}
\end{eqnarray}
where $M_{\mu}$ is the muon mass and we neglected the electron
mass. However the effects of a nonzero electron mass are as in the
Standard Model case \cite{Hagiwara:fs}. The parameter $\delta$ can
be estimated using the vacuum polarizations presented in the
previous section yielding:
\begin{equation}
\delta\approx\frac{g^2}{3(4\pi)^2}\frac{\mu_A^2}{M_W^2}\log\left(\frac{(2+\sqrt
3)\mu_A^2}{\Lambda^2}\right).
\end{equation}
By choosing the renormalization scale $\Lambda$ to be of the order
of the electroweak scale $\sim M_Z$ and $\mu_A\simeq 150$~GeV we
determine $\delta\simeq 0.007$. Precise and independent
measurements of $M_W$ and the muon decay rate may observe
deviations with respect to the Standard Model. Finally we expect
sizable corrections to the fermion dispersion relations in
eq.~(\ref{fermion-dr}) induced by the gauge boson exchanges. These
arise in the fermion vacuum polarization at the two loop level and
are expected to be of the order of ${g^2}\delta/{(4\pi)^2}$.

\section{Model II: Hidden Bose-Einstein Mechanism}

Another model was proposed in \cite{Sannino:2003ai}. Now the
Standard Model is extended by adding a new complex bosonic field
$\phi$ which is a singlet under all of the standard model charges.
In this way we also introduce an extra $U(1)$ symmetry. The most
general renormalizable potential involving the standard model
Higgs and the field $\phi$ respecting all of the symmetries at
hand is:
\begin{eqnarray}
V\left[\phi,M\right]&=&\frac{1}{2}\left(M^2_{H}-8
\hat{g}\,\left|\phi\right|^2\right){\rm
Tr}\left[M^{\dagger}M\right] + m^2|\phi|^2 \nonumber \\
&+&{\hat{\lambda}}\left|\phi\right|^4 + {\lambda} {\rm
Tr}\left[M^{\dagger}M\right]^2 \ ,
\end{eqnarray}
where $M\equiv\frac{1}{\sqrt{2}}\left(\sigma + i\,
\vec{\tau}\cdot\vec{\pi} \right)$. We have assumed the potential
to be minimized for a zero vacuum expectation values of the fields
and the couplings to be all positive. With this choice it is clear
that if $\phi$ acquires a non zero vacuum expectation value the
ordinary Higgs also acquires a negative mass square contribution.

We introduce a non zero background charge for the field $\phi$,
which modifies the $\phi$ kinetic term as follows
\cite{Sannino:2003mt, Haber:1981ts, Kapusta:aa}:
\begin{eqnarray}{\cal L}_{\rm charge}={\cal{D}}_{\mu}\phi^{\ast} {\cal{D}}^{\mu}\phi\ , \label{bose}
\end{eqnarray}
with
\begin{eqnarray}
{\cal{D}}_{\nu} \phi=\partial_{\nu}\phi - i {\cal{A}}_{\nu}\phi \
, \qquad {\cal{A}}_{\nu}=\mu \left(1,\vec{0}\right) \ ,
\label{covariant}
\end{eqnarray}
and $\mu$ is the associated chemical potential. Substituting
(\ref{covariant}) in ({\ref{bose}}) we have:
\begin{eqnarray}
{\cal L}_{\rm charge}=\partial_{\mu}\phi^{\ast}\partial^{\mu}\phi
+i\,\mu\left(\phi^{\ast}\partial_{0}\phi -
\partial_{0}\phi^{\ast}\phi\right) + \mu^2\,|\phi|^2\ . \label{charge}\end{eqnarray}
The introduction of the chemical potential has broken Lorentz
invariance $SO(1,3)$ to $SO(3)$ while providing a negative mass
squared contribution to the $\phi$ boson. When $\mu>m$ the
spontaneous breaking of the $U(1)$ invariance is a necessity. Once
the bosonic field has acquired a vacuum expectation value and for
$8\hat{g}\,\langle |\phi| \rangle^2 > M^2_H $ the Higgs field
condenses as well.

In this model the Bose-Einstein mechanism, although indirectly,
still triggers electroweak symmetry breaking while the effects of
Lorentz breaking induced by the presence of the chemical potential
are attenuated. The latter are controlled by the new coupling
constant $\hat{g}$ as well as the ordinary higher order
electroweak corrections. Our model potential is similar in spirit
to the one used in hybrid models of inflation \cite{Linde:1993cn}.

{}For the parameter values $M^2_H=m^2=0$ it was shown in
\cite{Sannino:2003ai} that the ground state of the theory has
\begin{eqnarray}  \quad \langle |\phi| \rangle^2 =
\frac{\mu^2}{2\hat{\lambda}} + {\cal O}(\epsilon^4) \ , \qquad
\langle \sigma \rangle ^2 =
\frac{\mu^2}{2\hat{\lambda}}\,\epsilon^2+ {\cal O}(\epsilon^6)
\nonumber  \ ,
\end{eqnarray}
where $\epsilon^2=2\hat{g}/\lambda\ll 1$. Assuming the new physics
scale associated to $\langle |\phi| \rangle$ to be within reach of
LHC, i.e., of the order of $1- 10$~TeV while the Higgs scale is
$\langle \sigma \rangle\simeq 250$~GeV we determine:
\begin{eqnarray}
\epsilon \simeq 0.25-0.025 \ .
\end{eqnarray}
All the corrections will be proportional to the fourth power of
$\epsilon$. We adopt the unitary gauge for the Higgs field and use
the polar decomposition for the field $\phi$:
\begin{eqnarray} \sigma =\langle \sigma \rangle + h \ , \quad
\phi=\frac{\left( \sqrt{2}\langle \phi \rangle + \psi
\right)}{\sqrt{2}} \,e^{i\frac{\eta}{\sqrt{2}\langle \phi
\rangle}} \ .
\end{eqnarray}
In the broken phase $h$ and $\psi$ are not mass eigenstates. These
are related to $h$ and $\psi$:
\begin{eqnarray}
h=\cos\theta \, \widetilde{h} - \sin\theta \, \widetilde{\psi} \
,\qquad \psi=\cos\theta\, \widetilde{\psi} + \sin\theta\,
\widetilde{h} \ .
\end{eqnarray}
Working out the details of the diagonalization
\cite{Sannino:2003ai}, one finds that the new neutral Higgs field
$\widetilde{h}$ due to the mixing with $\psi$ feels feebly but
directly the presence of the net background charge. The propagator
for $\widetilde{h}$ is:
\begin{eqnarray}
\frac{i}{p^2-m^2_{\widetilde{h}}-4 p^2_0 \mu^2 \,\sin^2\theta
\,{\cal F}\left[p,p_0 \right]} \ , \label{propagator}
\end{eqnarray}
with
\begin{eqnarray}
{\cal F}\left[p,p_0\right]=\frac{m^2_{\widetilde{\psi}}-
p^2}{\left(m^2_{\widetilde{\psi}}-
p^2\right)\,p^2+4p_0^2\,\mu^2\,\cos^2\theta} \ .
\label{propagator2}
\end{eqnarray}
We take $\psi$ to be heavy so that the only corrections will be
the ones induced by the small mixing between ${h}$ and $\psi$. The
function (\ref{propagator2}) for large $\mu$ is ${\cal F}\approx
1/(3p^2_0-{\bf {p}}^2)$ and resembles the pole associated to the
gapless state $\eta$. Again, if phenomenologically needed we can
give a small mass (with respect to the scale $\mu$) to $\eta$
\cite{Sannino:2003mt}. The rest of the standard model particles
are affected by the presence of a frame via weak radiative
corrections and these corrections can be computed as in
\cite{Sannino:2003mt}. The Higgs propagator here is more involved
than the one in \cite{Sannino:2003mt}. However the relevant point
is that the effects of the non standard dispersion relations in
the model II are damped with respect to the ones determined in
model I \cite{Sannino:2003mt} due to the presence of the
suppression factor $\sin^2\theta\approx \epsilon^6$ in the
$\widetilde{h}$ propagator. Indeed since
$m^2_{\widetilde{h}}\approx\epsilon^2 \mu^2$ the term in the
propagator bearing information of the explicit breaking of Lorentz
invariance due to the Bose-Einstein mechanism is down by
$\epsilon^4$ with respect to the mass term. So we have a large
suppression of Lorentz breaking effects induced by the presence of
a frame needed for the Bose-Einstein mechanism to take place.
Clearly also the dispersion relations of the ordinary Higgs are
modified only slightly with respect to the standard scenario. The
corrections due to the mixing with the $\psi$ field should also be
taken into account although in general they will be further
suppressed due to the assumed hierarchy $\mu\gg m_h$.

Let us roughly estimate, using the results of
\cite{Sannino:2003mt}, the size of the corrections to some
observables. We first consider the modification of the low energy
Fermi theory due to the new sector. Recall that the parameter
$\delta$ controlling the corrections in the low energy theory
(\ref{fermitheory}) was found to be $\delta \approx 0.007$ in the
direct Bose-Einstein model. {}For the hidden Bose-Einstein sector
this result, on general grounds, is further suppressed by a
multiplicative factor of the order $\epsilon^4$ yielding a new
$\delta$ of the order of $2.7\times 10^{-5} - 2.7 \times 10^{-9}$
for $\langle |\phi| \rangle \approx 1 - 10$~TeV.

The fermion sector also bears information of the direct or
indirect nature of the underlying Bose-Einstein phenomenon. We
demonstrated for the direct case in \cite{Sannino:2003mt} that the
one loop corrections to the fermion velocities due to the exchange
of the Higgs are tiny ($\simeq 10^{-15}$ for the electron). This
is due to the smallness of the associated Yukawa's couplings.
However we have also argued that the higher order corrections due
to the modified gauge boson dispersion relations may be relevant
and estimated in this case a correction of the order of
$\delta\,g^2/4\pi$ with $g$ the weak coupling constant. In the
hidden Bose-Einstein case also the corrections to this sector are
further suppressed by a factor of the order of $\epsilon^4$ with
respect to the findings in \cite{Sannino:2003mt}.

\section{Conclusions}

The Bose-Einstein condensation phenomenon has been proposed as
possible physical mechanism underlying the spontaneous symmetry
breaking of cold gauge theories. We have explicitly considered two
possible realizations of this picture. In model I
\cite{Sannino:2003mt} the Higgs field was assumed to carry global
and local symmetries and was identified with the Bose-Einstein
field. The effects of a non zero background charge were, in this
way, maximally felt in the Higgs sector and then communicated via
weak interactions to the other standard model particles. In model
II \cite{Sannino:2003ai} the Higgs mechanism was triggered by a
hidden Bose-Einstein sector. The relevant feature of this model is
that the effects of modified dispersion relations for the standard
model fields are strongly suppressed with respect to the ones in
model I. We have predicted the general form of the corrections and
estimated their size in both cases. The fermion sector of the
Standard Model, for example, is affected via higher order
corrections leading to the appearance of modified dispersion
relations of the type $E^2 = v_{f}^2\,p^2+m_f^2$. The deviation
with respect to the speed of light for the fermions is small when
considering the direct Bose-Einstein mechanism
\cite{Sannino:2003mt} while it is further suppressed by a factor
$\epsilon^4$ in the model presented here.

We emphasize that the form of the corrections, induced by the
modified propagators in Eqs. (\ref{propagator}) and
(\ref{propagator2}), are general and dictated solely by the
Bose-Einstein nature of our mechanism. The strength of the
coupling between the standard model fields and the Bose-Einstein
field is measured by the parameter $\epsilon$ which enters in some
of the physical observables. Experiments can be used to constrain
the possible values of $\epsilon$.

An interesting consequence of the direct (i.e. model I)
Bose-Einstein condensation mechanism is that the mass $\Delta$ for
the neutral Higgs $h$ is predicted to be larger than the one in
vacuum i.e. $M_h^2=2\left(\mu^2_A - m^2\right)$. More specifically
$\Delta \geq \sqrt{3} M_h$. If, for example, the in vacuum Higgs
mass is chosen to be about $90$~GeV the Bose Einstein mechanism
yields a mass larger than $155$~GeV.

Spontaneous breaking of a gauge theory via Bose-Einstein
condensation necessarily introduces Lorentz breaking since a frame
must be specified differentiating time from space. We recall that
the issue of Lorentz breaking has recently attracted much
theoretical
\cite{Kostelecky:2002hh,Carroll:2001ws,{Amelino-Camelia:2002dx}}
and experimental attention \cite{Carroll:vb}. In the Bose-Einstein
case the underlying gravitational theory is not the cause of
Lorentz breaking which is instead due to having immersed the
theory in a background charge.

Modified dispersion relations for the standard model particles
have also cosmological consequences since now one cannot simply
assume the velocity of light as the common velocity for all of the
massless elementary particles. In fact different particles will
have, in general, different dispersion relations and hence speed.
This is relevant, for example, when observing neutrinos from
distant sources. Using the present model as guide for the
structure of the corrections experiments can test the
Bose-Einstein mechanism as possible source of electroweak symmetry
breaking.

\newpage
\acknowledgments We would like to thank E. Ferrer and V.~de la
Incera for relevant discussions.

\end{document}